# Interesting history effect in the magnetotransport properies of $La_{0.55}Ho_{0.15}Sr_{0.3}MnO_z$ films on $LaAlO_3$

P Raychaudhuri[a]Å , A E P de Araujo[b], F L A Machado[b], A K Nigam[a], R Pinto[a]

[a)] ***Tata Institute of Fundamental Research, Homi Bhabha Rd., Mumbai 400005, India.***

[b)] ***Department of Physics, Universidade Federal de Pernambuco, Recife, Brazil.***

***Abstract:*** We report the magnetoresistance measurements of a highly oriented $La_{0.55}Ho_{0.15}Sr_{0.3}MnO_z$ film on $LaAlO_3$ substrate. The film has a metal-insulator transition around 200 K and shows pronounced thermomagnetic history effect in its transport properties below 45 K at 7 Tesla. The irreversibility temperature shifts to lower temperatures when the field is reduced. This behaviour is opposite to the thermomagnetic history effect observed in the magnetisation. At 2.8 K one also observes a significant hysteresis in the resistance versus field curve. We propose a qualitative explanation for these observations.

---

Å e-mail:prat@tifr.res.in

# I. **Introduction**

There has been a recent surge of interest in colossal magnetoresistive (CMR) hole doped rare-earth manganites of the form $R_{1-x}A_xMnO_3$ ( R = rare-earth, A = bivalent cation) due to the plethora of new and interesting behaviours observed in these compounds. For $x \geq 0.2$ the material shows a ferromagnetic transition with a metal-insulator transition (MIT) and very large negative magnetoresistance $(MR\sim(\rho(0)-\rho(H))/\rho(0))$ close to the ferromagnetic transition temperature $T_c$ [1,2,3]. The basic underlying physics of these compound is described by the Zener double exchange [4] mechanism by which an electron can hop between two neighbouring $Mn^{3+}/Mn^{4+}$ ions via the intermediate $O^{2-}$ ion. Though this simple mechanism explains many of the observed features in these compounds several other features remain unexplained within the realm of this model. These systems therefore call for study in greater detail before their properties are fully understood.

Substitution of other rare-earth at the La has revealed many interesting properties of these materials. It has been observed that when the size of the rare-earth is decreased the metal-insulator transition and ferromagnetic transition shift to lower temperatures [ ]. At small enough size the system is driven into a spin glass state with very large magnetoresistance below the spin glass transition temperature [ ]. This phenomenon has been explained in terms of the competition between the antiferromagnetic superexchange and the ferromagnetic double exchange interaction in the material. It is thus interesting to probe the properties of this material in the epitaxial thin film form.

In this paper we report a new feature in thin films of the CMR manganite $La_{0.55}Ho_{0.15}Sr_{0.3}MnO_z$, namely, an unusual magnetic history effect in field cooled (FC) and zero field cooled (ZFC) resistance of the sample. In contrary to ones intuition the irreversibility shifts to lower temperatures at lower values of field and is totally absent at 1 kOe. We propose a qualitative explanation for these observations.

## II. **Experimental Details**

Thin films of $La_{0.55}Ho_{0.15}Sr_{0.3}MnO_z$ were grown on $LaAlO_3$ substrate from a stoichiometric target by pulsed laser deposition in oxygen ambient. The substrate temperature was kept at $760^0$C and the ambient oxygen pressure was kept at 400 mTorr. X-ray diffraction (XRD) $\theta$-$2\theta$ showed very good orientation of the rhombohedral unit cell with the (110) direction perpendicular to the film surface. Figure 1 shows X-ray $\phi$-scan of the (10-1) family of peaks carried out on a 4-circle goniometer. The sharp peaks at $90^0$ intervals show very good in-plane orientation of the film. The FC and ZFC resistance and the magnetoresistance isotherms were measured using conventional 4-probe technique employing a high field superconducting magnet generating a field up to 8 Tesla. The field cooling and zero field cooling were done by heating the sample to a temperature of 150 K.

## III. **Results and Discussion**

Figure 2(a) shows the resistance versus temperature (R-T) in zero field and at a field of 7 Tesla. The resistance in zero field shows a MIT around 200 K ($T_p$) which increases to 250 K in the presence of the field. However, it is the interesting to note that the film shows a minimum in the resistance at low temperature. This minimum is absent in $La_{0.7}Sr_{0.3}MnO_3$ film (inset fig.2(d)). This minimum was observed in all our Ho doped films and has been reported earlier [11]. Though the origin of this minimum in a ferromagnetic sample is not clear it seems to be associated with the magnetic scattering of Ho in the sample.  The maximum in the MR at 7 Tesla (also shown in the same figure) occurs slightly below the metal-insulator transition temperature. The interesting feature is that the MR does not drop very rapidly below the metal-insulator transition and remains significant even at the lowest temperatures. This feature which is interesting from the application point of view is not observed in epitaxial films of $La_{0.7}Ca_{0.3}MnO_3$ [12]. In the case of polycrystalline samples this behaviour comes from spin polarised tunnelling at the grain boundaries [13,14], where this effect gives rise to a sharp drop in resistance below the technical saturation field of the ferromagnet. Such a drop is absent here. To understand the possible reason for this behaviour we have compared the R-T curve of an epitaxial $La_{0.7}Ca_{0.3}MnO_3$ film (fig. 2(d)) with the one of $La_{0.55}Ho_{0.15}Sr_{0.3}MnO_3$. We observe that the $La_{0.7}Ca_{0.3}MnO_3$ film has a much sharper metal-insulator transition than the $La_{0.55}Ho_{0.15}Sr_{0.3}MnO_3$ film. It is thus likely that the substitutional disorder arising due to the presence of three different ionic radii at the rare-earth site broadens the ferromagnetic transition giving rise to large MR below the metal-insulator transition. This will be discussed further when discussing the thermomagnetic history effect. Figures 2(b) and 2(c) show the 4 quadrant the resistance versus field (R-H) curve at 2.8 K and 4.5 K respectively. These measurements were done after zero field cooling of the sample. The interesting point to note is that there is a large hysteresis between the initial increasing field curve and the decreasing field curve. Under subsequent field cycling the hysteresis greatly decreases but does not totally vanish. However one cannot obtain the initial state of the system. This shows that the initial state of the system is different from the state obtained after field cycling. The hysteresis of R-H curve persists up to 25 K (*inset* of figure 2 (b)).

Figure 3(a)-(e) shows the resistance versus temperature under FC and ZFC in various cooling fields. At 8 Tesla (fig. 3(e)) FC and ZFC curve bifurcates at around 55

K. At 0.1 Tesla there is no observable difference between FC and ZFC resistance. This unusual behaviour excludes the possibility of a spin glass like freezing or domain wall pinning playing a role in the transport since in both these cases the irreversibility is known to increase in lower fields. FC and ZFC resistance measurements were done at various other intermediate fields. Figure 3(f) shows the variation of the irreversibility temperature ($T_{irr}$, defined as the temperature where the difference between FC and ZFC resistance starts) at different fields. The irreversibility temperature decreases slowly below 8 Tesla up to 3 Tesla but then drops off rapidly. It is thus clear from these data that the irreversibility in FC and ZFC magnetisation in these materials is not of the same origin as that conventionally attributed to the irreversibility in the magnetisation.

In figure 3(g) we show the R-T curve for the sample cooled in different field. The sample was cooled in field to the lowest temperatures and the field was switched off. The data was taken while warming up in zero field. The resistance value is lower at low temperatures when the sample is cooled at higher fields again showing the thermomagnetic history dependence of resistance in this material. The minimum shifts to lower values of temperature when the sample is cooled in higher field showing that the resistance has a relaxation when the field is switched off.

In order to understand the above mentioned phenomenon one has to imagine the $La_{0.55}Ho_{0.15}Sr_{0.3}MnO_z$ film to have some magnetic inhomogeneity. One possibility is that conducting ferromagnetic regions are separated by frozen in regions with very small volume fraction. One supportive evidence for this is the very small hysteresis in magnetisation one observes in these samples particularly at high fields. Under the application of a magnetic field the spins in these frozen in regions will tend to align parallel to the magnetic field giving a drop in the resistance. This is likely to be the origin of the large MR observed at temperatures much below the metal-insulator transition. A similar situation is believed to exist in $Nd_{0.7}Sr_{0.3}MnO_z$ thin films [15]. It should be noted here that the field cooling and zero field cooling in these samples were done from 150 K and not above $T_p$. It might thus be possible that at low fields the frozen in configuration remains unaffected by the application of field. It is only at high fields that one obtains a significant difference between FC and ZFC resistance.

In summary, we report in this paper an unusual property of $La_{0.55}Ho_{0.15}Sr_{0.3}MnO_3$ epitaxial thin film, namely, the difference between the field cooled and zero field cooled resistance. The strong irreversibility in the resistance at high fields indicates that the irreversibility mechanism in the electronic transport is different from the conventional mechanism in magnetisation. We have proposed a possible explanation for these observations which, however, have to be confirmed through further experiments.

## Figure Captions

Figure 1. $\phi$–scan of the (1 0 -1) family of peaks showing very good in plane orientation of the film.

Figure 2. (a) Resistance versus temperature in zero field and at 7 Tesla. The solid line shows the MR as a function of temperature. (b) Resistance versus field (R-H) at 2.8 K. (c) R-H at 4.5 K; the *inset* shows R-H at 25 K. (d) Resistance versus temperature for $La_{0.7}Ca_{0.3}MnO_3$; the inset shows the low resistance versus temperature for $La_{0.7}Sr_{0.3}MnO_3$ at low temperature.

Figure 3. (a)-(e) Field cooled and zero field cooled resistance as a function of temperature for different cooling field; (f) $T_{irr}$ as a function of cooling field; (g) resistance versus temperature taken while warming in zero field after cooling the sample in various fields.

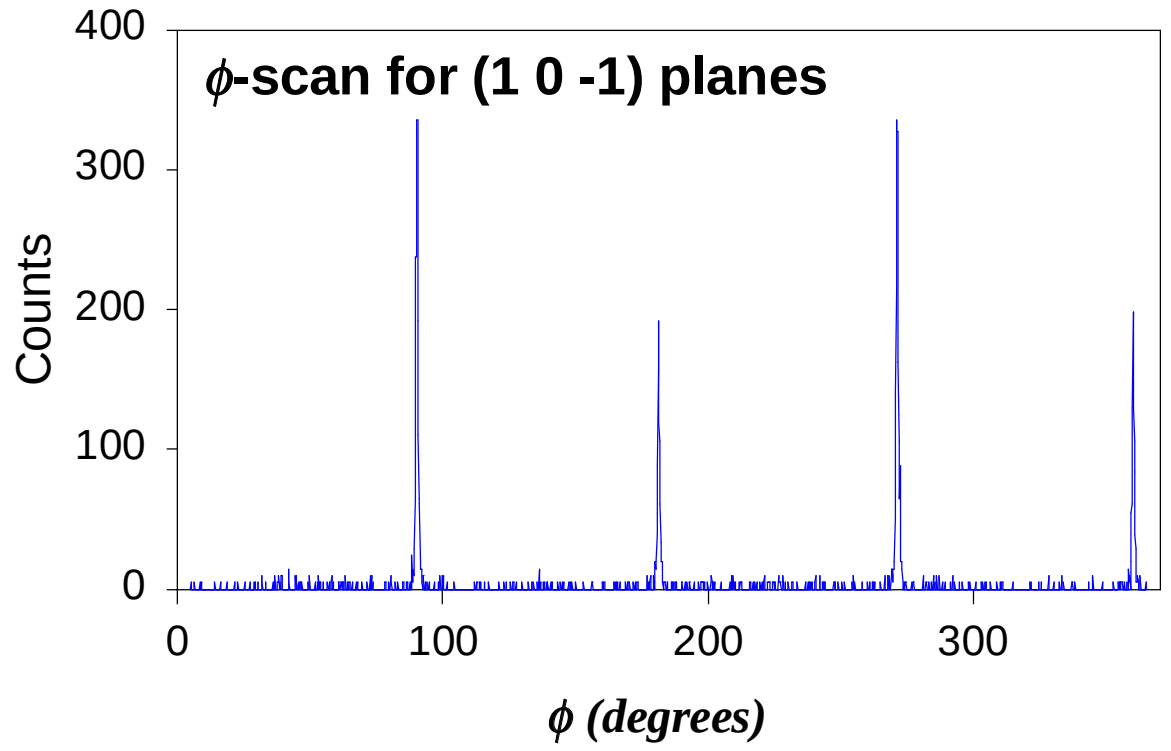

**Figure 1 (P Raychaudhuri et al)**

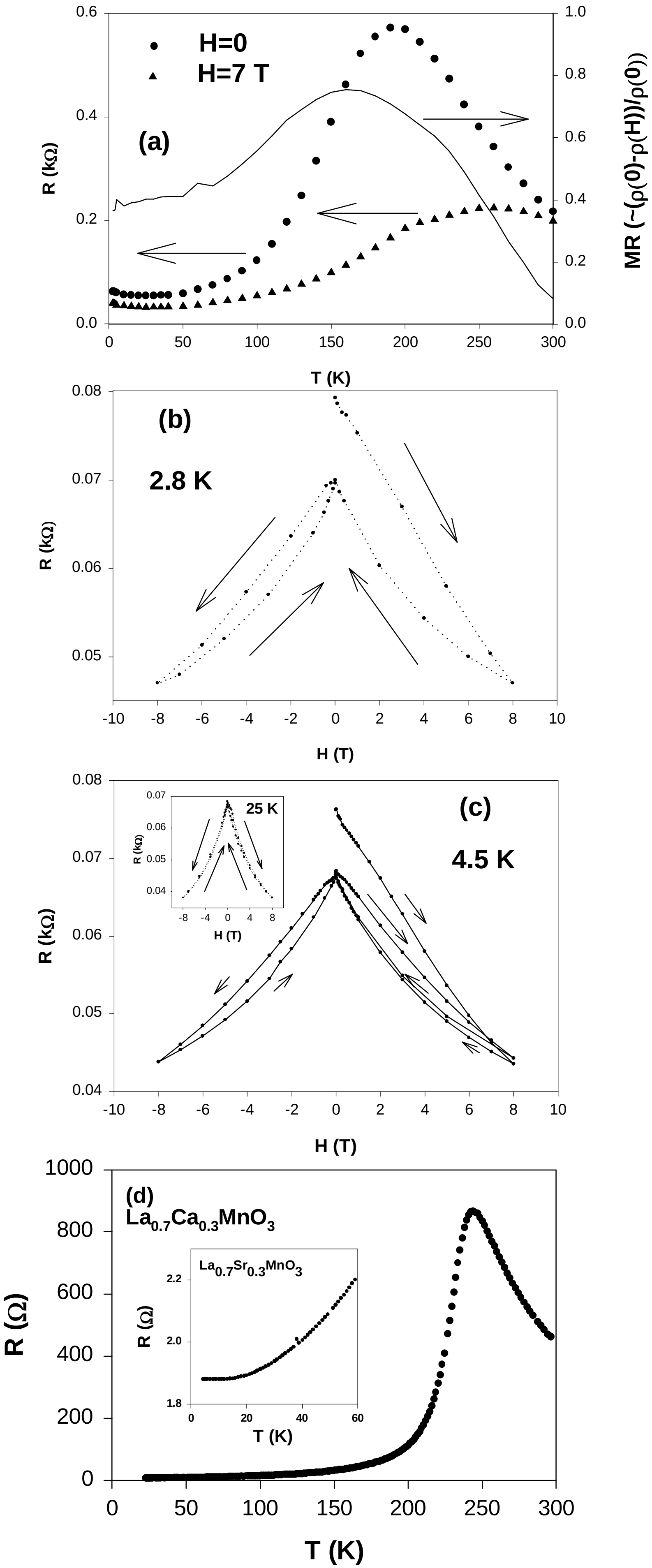

Figure 2 (P Raychaudhuri et al)

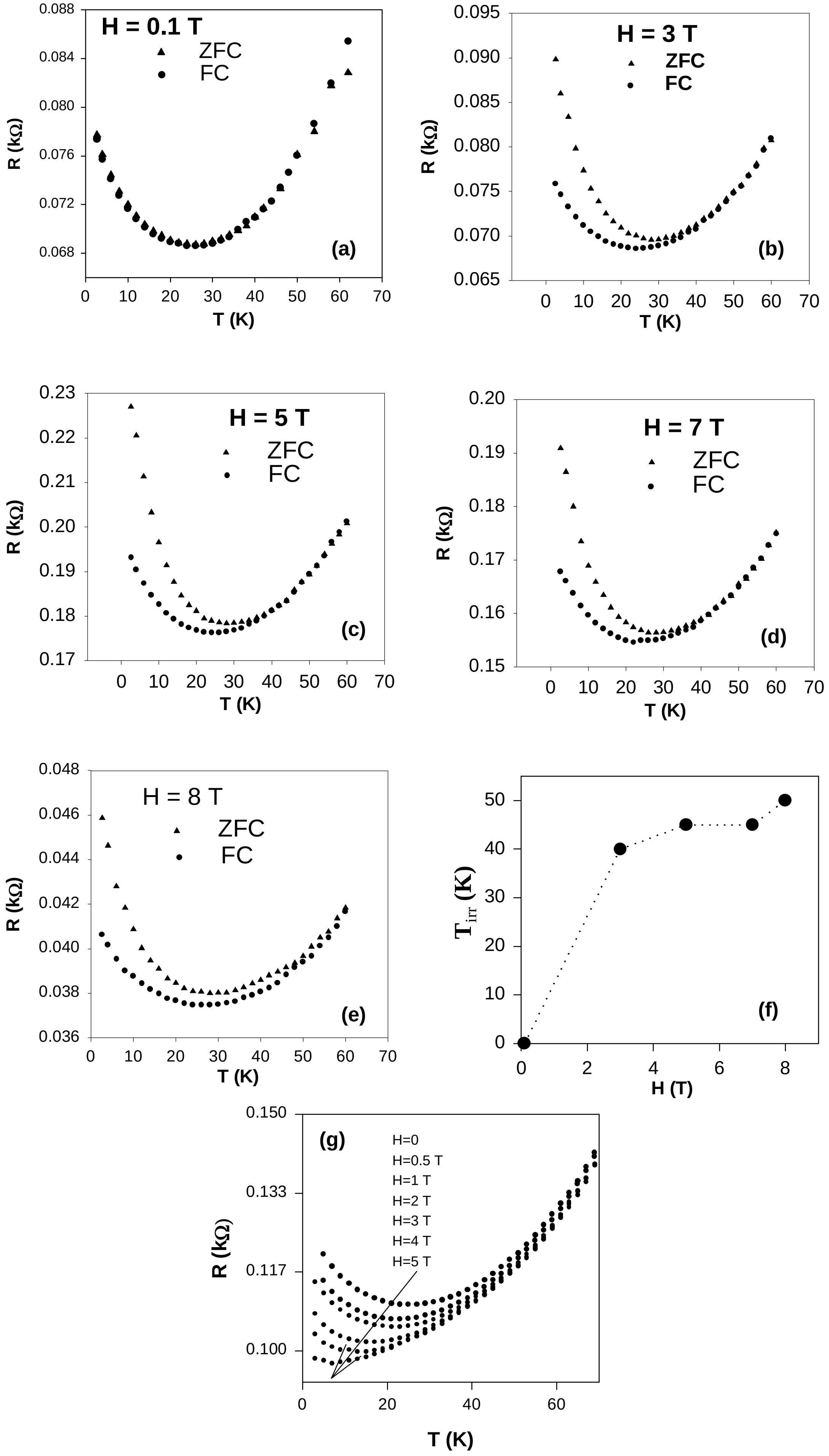

Figure 3 (P Raychaudhuri et al)